\documentclass[10pt,twocolumn,letterpaper]{article}

\usepackage[cvpr]{cvpr}      

\usepackage{tikz}
\usepackage{xcolor}
\usepackage{xspace}

\newcommand{\highlight}[2]{%
    \definecolor{tempcolor}{HTML}{#1}%
    \tikz[baseline=(X.base)] 
      \node[
        rectangle, 
        fill=tempcolor, 
        anchor=base, 
        inner sep=2pt,
        minimum width=1.2cm,  %
        minimum height=0.5cm, %
        rounded corners=3pt
      ] (X) {#2};%
}
\newcommand{\highlightbox}[2]{%
  \definecolor{tempcolor}{HTML}{#1}%
  \tikz[baseline=-0.5ex] 
    \node[
      rectangle, 
      fill=tempcolor, 
      anchor=base, 
      inner sep=2pt,
      minimum width=0.5cm,
      minimum height=0.5cm,
      rounded corners=3pt
    ] {#2};%
}

\newcommand{\btoken}{\highlight{D4E8D3}{\small \textbf{$<$group$>$}}\xspace}
\newcommand{\bostoken}{\highlight{F2F2F2}{\small \textbf{$<$BOS$>$}\xspace}}
\newcommand{\clstoken}{\highlightbox{FDE5D6}{\small $\mathbf{F}_{\text{cls}}$}\xspace}
\newcommand{\ttoken}{\highlightbox{E6F1E2}{\small $\hat{\mathbf{T}}$}\xspace}
\newcommand{\ptoken}{\highlight{D5E9F7}{\small density}\xspace}
\newcommand{\tboken}{\highlightbox{E6F1E2}{\small $\mathbf{T}$}\xspace}

\definecolor{cvprblue}{rgb}{0.21,0.49,0.74}
\usepackage[pagebackref,breaklinks,colorlinks,allcolors=cvprblue]{hyperref}

\usepackage{multirow}
\usepackage[table,xcdraw]{xcolor}
\usepackage{colortbl}

\def\paperID{568} 
\def\confName{CVPR}
\def\confYear{2026}

\title{Animator-Centric Skeleton Generation on Objects with Fine-Grained Details}

\author{
Mingze Sun$^{1,2}$\thanks{Equal Contribution.}
\hspace{0.8em}
Cheng Zeng$^{1}$\footnotemark[1] \hspace{0.8em}
Jiansong Pei$^{1}$ \hspace{0.8em}
Junhao Chen$^{1}$ \hspace{0.8em}
Chaoyue Song$^{3}$ \hspace{0.8em}
Shaohui Wang$^{2}$ \\
Tianyuan Chang$^{2}$ \hspace{0.8em}
Bin Huang$^{2}$ \hspace{0.8em}
Zijiao Zeng$^{2}$\footnotemark[2] \hspace{0.8em}
Ruqi Huang$^{1}$\thanks{Corresponding Author.} \\
$^{1}$Tsinghua Shenzhen International Graduate School, China $^{2}$Tencent VISVISE\\
$^{3}$Nanyang Technological University
}
\begin{document}

\twocolumn[{

\renewcommand\twocolumn[1][]{#1}
\maketitle

\begin{center}
    \captionsetup{type=figure}
    \centerline{\includegraphics[width=\linewidth]{figs/teaser.png}}
    \caption{
    We propose an automatic and controllable skeleton generation framework. Given an input mesh, our method generates fine-grained skeletal structures—including detailed human skirts and wide sleeves, as well as reins for horses—providing a strong foundation for producing high-quality animations.
    }
    \label{fig:teaser}

\end{center}

}]
\begingroup
\renewcommand\thefootnote{}
\footnote{$^*$ Equal Contribution. $^\dagger$ Corresponding Author.}
\addtocounter{footnote}{-1}
\endgroup

\begin{abstract}

Skeleton generation is essential for animating 3D assets, but current deep learning methods remain limited: they cannot handle the growing structural complexity of modern models and offer minimal controllability, creating a major bottleneck for real-world animation workflows.
To address this, we propose an animator-centric SG framework that achieves high-quality skeleton prediction on complex inputs while providing intuitive control handles. Our contributions are threefold. First, we curate a large-scale dataset of 82,633 rigged meshes with diverse and complicated structures. 
Second, we introduce a novel \emph{semantic-aware tokenization} scheme for auto-regressive modeling. 
This scheme effectively complements purely geometric prior methods by subdividing bones into semantically meaningful groups, thereby enhancing robustness to structural complexity and enabling a key control mechanism. 
Third, we design a \emph{learnable density interval module} that allows animators to exert soft, direct control over bone density.
Extensive experiments demonstrate that our framework not only generates high-quality skeletons for challenging inputs but also successfully fulfills two critical requirements from professional animators. 

\end{abstract}    
\section{Introduction}
\label{sec:intro}

\emph{La vie est dans le mouvement. -- Voltaire}

3D animators breathe a spirit into digital assets by driving a \emph{static} object into a sequence of its \emph{dynamic} variants. 
During this magic procedure, skeleton generation (SG) plays a fundamental role -- the skeleton not only serves as an effective simplification of the asset, but also as a powerful and initial editing handle for animators. 

Due to the labor-intensive and time-consuming nature of hand-crafting SG, researchers have made efforts towards the automation of SG. 
Early approaches~\cite{xu2020rignet, ma2023tarig} typically cast SG as a geometric optimization problem, while being axiomatic, it is non-trivial to inject either animators' expertise or semantic understanding. 
A more recent trend~\cite{song2025magicarticulate, sun2025armo, song2025puppeteer} is to approach in a data-driven manner. 
In a nutshell, deep neural networks are trained to align the \emph{geometric} encoding of an input object and the regarding manual skeleton label. 

Yet, we argue that there are two major bottlenecks shared by the prior automatic SG frameworks. 
First of all, the pursue of more and more realistic animation is persistent~ consistent, and long-standing\footnote{For instance, the character of Super Mario has evolved significantly from the 1980s (on Nintendo FC) to the 2020s (on Nintendo Switch).}. 
Beyond that, the recent advances of 3D generative models~\cite{li2025droplet3d, chen2025ultra3d, lai2025hunyuan3d, hunyuan3d2025hunyuan3d} have further boosted digital content production -- high-quality 3D assets with complicated structures can now be created at large scale, high speed, and low cost. 
It is then crucial to develop an SG framework capable of dealing with rich structures from various sources composed on a single object (\emph{e.g., }the character with a complex hairstyle and clothing shown in Fig.~\ref{fig:teaser}(a)). 
Unfortunately, the existing methods typically treat the 3D object as a whole and rely heavily on geometric encoding/prior, making them challenging to adapt the increasing structural complexity. 

Second, to the best of our knowledge, crafting motions on arbitrary skeletons remains heavily dependent on manual efforts. 
Thus, it is indeed crucial to allow animators to gain as much control as possible over the SG procedure. 
However, the existing methods, either axiomatic or learning-based, tend to be \emph{end-to-end without conditional control}~\footnote{One exception is RigNet~\cite{xu2020rignet}, with which users can tentatively increase/decrease bone number by tuning a certain threshold. Yet, the control is rather qualitative and heuristic.}. 
In other words, most of the time, animators can only perform non-trivial post-processing of the SG results, which severely hinders flexibility and efficiency in the subsequent animation stages. 

Motivated by the above, our goal is to establish an \emph{animator-centric} SG framework, which not only achieves high-quality skeleton prediction on challenging input with complicated structures, but also offers control handles to animators for easing customization within the animation pipeline. 
In particular, based on communications with animators from industry, we prioritize the following two requirements: (\textbf{R1}) Animators desire to \emph{designate} their crafted skeleton at a coarse level or some local region of interest; (\textbf{R2}) Animators appreciate a more \emph{direct and explicit} control over bone density. 

Targeting at the above, we present a large-scale auto-regressive model for skeleton generation, to which we devote efforts from the following two perspectives. 
\textbf{Data Preparation:} We curate a large-scale dataset of $82,633$ rigged meshes, which spans a range of categories and demonstrates varying structural complexity -- bone number ranges from $5$ to $400$. 
We highlight our dataset stands in stark contrast to prior ones such as ModelsResource~\cite{xu2020rignet} and Articulation-XL~\cite{song2025magicarticulate}, which are dominated by annotated skeletons of simple structures featured by low bone number (See Sec.~\ref{sec:dataset} for more details). 
\textbf{Model Design:} Though auto-regressive modeling is popular in rigging research~\cite{song2025magicarticulate, sun2025armo, song2025puppeteer}, their tokenization typically follows Breadth-First Search (BFS) order to the encoder skeleton, which is purely geometric. 
Such a naive scheme can suffer from increasing structural complexity due to geometric ambiguity 
UniRig considers a group-based decomposition strategy, but it relies on manually defined priors and cannot reliably preserve the local geometric topology of the skeleton -- for instance, in Fig.~\ref{fig:teaser}(b), UniRig incorrectly connects the rein to the horse's neck. 
To this end, we propose a novel semantic-aware tokenization scheme, which not only nicely complements the geometric tokenization but also helps to reduce structural complexity by subdividing bones into smaller groups of similar semantic meaning. 
Somewhat surprisingly, we further observe that the semantic-aware tokenization enables users to inject their own crafted coarse skeleton into the auto-regressive SG results, fulfilling (\textbf{R1}) above (see Sec.~\ref{sec:semantic} for details). 
On the other hand, regarding (\textbf{R2}), we propose a learnable density interval module, which produces \emph{density-aware} tokens into our auto-regressive model. 
In particular, instead of posing hard constraints on the exact bone number, our novel design softly encourages that of output skeleton falls into user-specified interval. 

We evaluate our SG model, both quantitatively and qualitatively, on our curated dataset, demonstrating that our approach effectively addresses the above challenges and achieves high-quality, controllable skeleton generation.

\section{Realated work}

\begin{figure*}[!t]
  \centering
  \includegraphics[width=\linewidth]{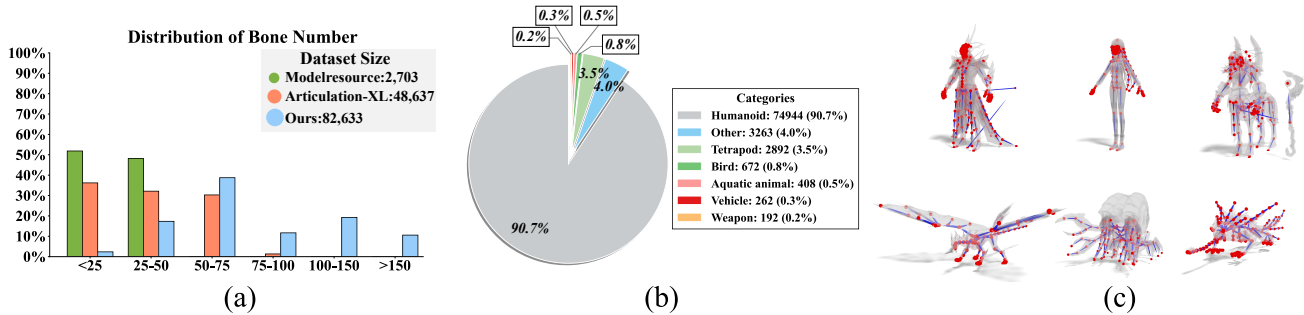}
    \caption{We compare the bone numbers distribution of our dataset with that of existing open-source datasets. (a) The dataset size and distribution of bone number across different datasets show that our dataset contains a wider range of articulated structures compared to ModelResource and Articulation-XL. (b) The category composition of our dataset, dominated by humanoid models, but also includes diverse non-humanoid types such as tetrapods, birds, aquatic animals, vehicles, and weapons. (c) Representative samples from various categories, visualized with skeletons and 3D meshes.
}
  \label{fig:dataset}
\end{figure*}

\subsection{Auto-regressive Model}
Auto-regressive (AR) modeling has achieved strong results across image, video, and 3D generation. Early works such as ImageGPT~\cite{Mark_Chen2020Imagegpt} and MaskGIT~\cite{chang2022maskgit} show that AR token prediction effectively captures spatial dependencies for high-fidelity image synthesis, with later models like Lumina-mGPT 2.0~\cite{xin2025lumina} scaling decoder-only architectures even further. In video generation, VideoGPT~\cite{Yan2021videogpt} and VideoPoet~\cite{Dan2023VideoPoet} extend this paradigm temporally to produce coherent, semantically rich videos.

Recently, AR modeling has been applied to 3D meshes. MeshGPT~\cite{siddiqui2024meshgpt} discretizes mesh patches into VQ-VAE codes and predicts them auto-regressively; MeshXL~\cite{chen2024meshxl} combines coordinate embeddings with neural decoders for meshes; and MeshAnything~\cite{chen2024meshanything} introduces conditional decoding and improved tokenization for scalable mesh synthesis. 
These decoder-only AR frameworks highlight the effectiveness of auto-regression for structured 3D generation, motivating its use in skeleton generation as well.

\subsection{Skeleton generation}
Early research on automatic skeleton generation leverages template input~\cite{baran2007automatic, li2021learning}.
Modern deep learning methods, such as RigNet~\cite{xu2020rignet}, pioneers an end-to-end neural skeleton generation pipeline that predicts joint locations via a geometry-aware graph network and estimates bone connectivity probabilities through a dedicated BoneNet. 
Building upon this framework, TARig~\cite{ma2023tarig} further improves skeleton generation quality for humanoid characters, while~\cite{guo2025make, chu2025humanrig} enhance skeleton generation performance for arbitrary poses and more heterogeneous characters.
However, regression-based approaches inherently limit the model’s generalization capability.

With the rapid progress of 3D generative modeling, recent works have begun to apply 3D generation frameworks to skeleton generation.
DRiVE~\cite{sun2025drive} first introduces a point-cloud diffusion model to generate joint positions, enabling the synthesis of skeletons for humanoid characters with clothing and hair.
However, the diffusion model can only produce joint positions.
MagicArticulate~\cite{song2025magicarticulate}
represents each bone as a token that jointly encodes parent–child geometry and semantic class. 
This formulation enables implicit connectivity learning and controllable structure synthesis but incurs redundancy and spatial ordering ambiguity. 
Concurrently, UniRig~\cite{zhang2025one} adopts a skeleton tree token strategy, but it relies on manually defined bone ordering and lacks automated semantic understanding of skeletal structures, which limits its generality and scalability.
Subsequent works, such as~\cite{liu2025riganything, sun2025armo, guo2025auto}, further improve skeleton prediction accuracy by adopting a two-stage paradigm and the RigFormer architecture, respectively.
Puppeteer~\cite{song2025puppeteer} redesigns the representation into joint-based tokenization with explicit parent indices and hierarchical breadth-first search (BFS) ordering, eliminating redundancy and stabilizing connectivity. 
However, these methods overlook the rich semantic structure of skeletons, making it difficult to generate complex and application-oriented skeletons that align with real-world requirements.
We propose a skeleton semantic understanding model and, based on it, design a semantic-based tokenization strategy that enables the auto-regressive model to generate high-quality and semantically coherent skeletons.

\section{Dataset Curation}\label{sec:dataset}
\begin{figure*}[h!]
  \centering
  \includegraphics[width=\linewidth]{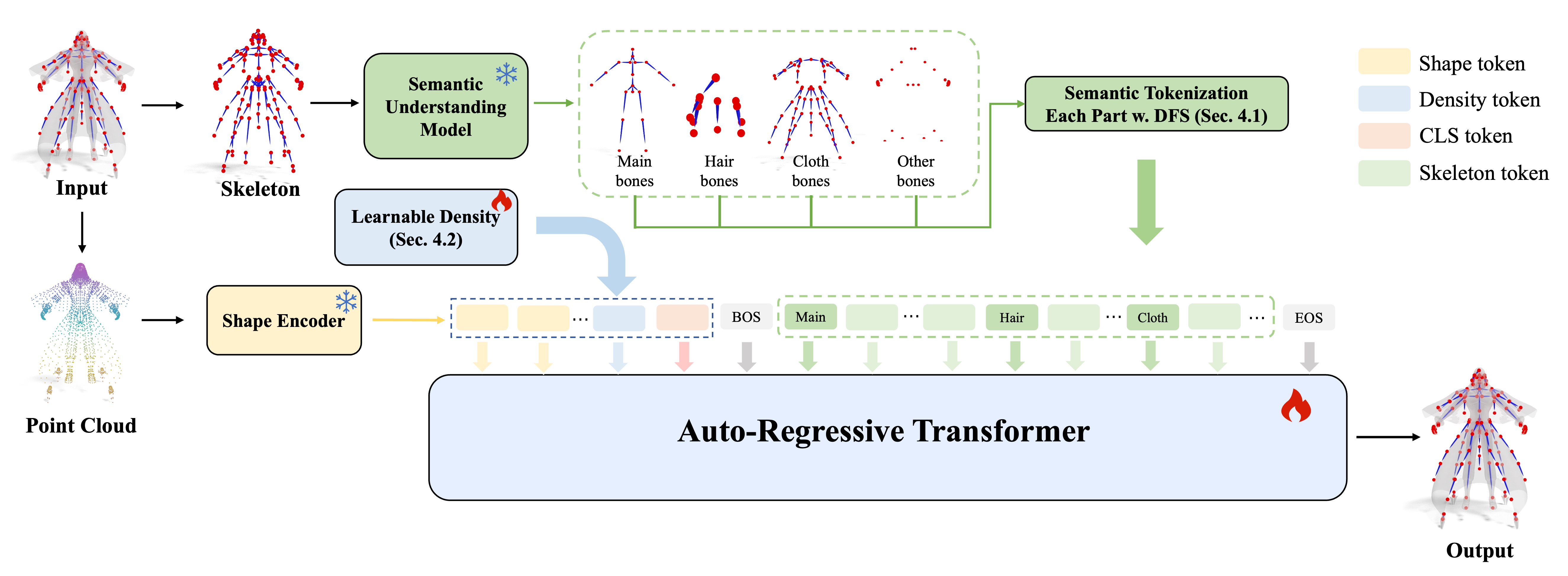}
    \caption{
   \textbf{The overall pipeline of our framework. }Given an input, the model first extracts geometric embeddings through a shape encoder. We introduce semantic-based skeleton tokenization through a semantic understanding model. A Learnable density token and a CLS token are added to realize controllable skeleton generation.}
  \label{fig:pipeline}
\end{figure*}

We collect over $150,000$ rigged 3D models from online sources, each containing rich rigging information.
To ensure data reliability and consistency, we set up a filtering pipeline to guarantee that each skeleton is well-aligned with its corresponding mesh, and that both joint positions and connectivity structures are accurate. 
Specifically, the pipeline is designed with the following principles. The technical details of filtering can be found in the Supp. Mat.

\begin{enumerate}
\item We ensure that the terminal joint of each bone chain lies within the reasonable geometric range of its corresponding mesh-connected component, thereby removing samples with drifting or penetrating skeletons;

\item We enforce that the skeleton hierarchy forms a single connected tree, eliminating models with multiple disconnected sub-trees or cyclic connections;

\item We discard samples with fewer than $5$ joints to maintain a minimum level of structural complexity for model training.

\end{enumerate}

After filtering, our final dataset consists of $82,633$ high-quality instance with bone number ranging from $5$ to $400$ -- the coarse distribution is shown in Fig.~\ref{fig:dataset}(a). 
Our dataset spans multiple categories, including humanoid, tetrapod, bird, aquatic, weapon, and vehicle, as illustrated in Fig.~\ref{fig:dataset}(b).
Compared to previous public rigging datasets~\cite{zhang2025one, song2025magicarticulate, sun2025armo}, in which the number of bones typically remains below $200$, our dataset exhibits a substantially higher structural complexity and greater category diversity.
We further perform stratified sampling across both categories and joint-count ranges, using $81,142$ shapes for training and $1,491$ shapes for testing, ensuring that the training and testing sets share similar distribution characteristics.

\section{Methodology}

We formulate skeleton generation as a conditional auto-regressive problem. 
Given an input mesh $\mathbf{M}$, our goal is to predict its corresponding skeleton $\mathbf{S}$, which consists of joint positions $\mathbf{J} \in \mathbb{R}^{k \times 3}$ and bone connections $\mathbf{B} \in \mathbb{R}^{b \times 2}$. 
In contrast to the recent works on auto-regressive rigging~\cite{song2025magicarticulate, sun2025armo, song2025puppeteer}, which depend on purely geometric tokenization, we introduce two novel tokenization schemes, which coordinate to address the challenges of 1) learning complicated skeleton structure and 2) allowing animators to inject their input (\emph{e.g.,} coarse main bones) or density preference into the generation results. 
In particular, we present our semantic-aware tokenization scheme in Sec.~\ref{sec:semantic}, and our learnable density control tokenization scheme in Sec.~\ref{sec:LDT}. 
Finally, we wrap up our tokenization strategy and further incorporate geometric information to finalize our training procedure in Sec.~\ref{sec:gpt}. 

\subsection{Semantic-Aware Tokenization}\label{sec:semantic} 
Note that the recent auto-regressive rigging frameworks~\cite{song2025magicarticulate, sun2025armo, song2025puppeteer} commonly adopt BFS order to encode the global skeleton topology, while being straightforward, this scheme is insufficient to handle geometric ambiguity in objects of increasingly complicated structure. 
To this end, we advocate to incorporate \emph{semantic} information for compensation.

Inspecting our dataset, we identify that the objects of high structural complexity mostly belong to the category of \texttt{humanoid} and \texttt{tetrapods}. 
In fact, the two categories also dominate our dataset in general -- they occupy nearly $95\%$ of data instances. 
For the sake of simplicity and efficiency, we design, train, and apply our semantic-aware tokenization only on those. 
Regarding the rest, we perform the Depth-First Search (DFS) algorithm to represent the skeletons as compact subsequences.

In the following, we start by pre-training a semantic understanding model, and then describe how to generate tokens upon it. 

\noindent\textbf{Semantic understanding model:} We first manually annotated $10,000$ instances with fine-grained semantic labels, which are sampled from humanoid characters and quadruped animals.
For humanoid skeletons, we define $29$ semantic categories: a) Main bones, such as head, shoulder, arms, torso, and legs; b) Auxiliary bones, such as hair, skirts, ribbons, and backpacks. 
Similarly, for tetrapods data, we categorize the skeletons into two major groups — main bones and auxiliary bones — resulting in $31$ subcategories in total.
The auxiliary bones include semantically meaningful structures such as fins, horns, and wings. 
Based on the annotated subset, we train a skeleton semantic prediction model to infer semantic labels for each joint.
The model takes as input the normalized joint positions along with the undirected graph representing the skeletal topology, enabling it to capture both geometric and structural relationships.
We adopt a GraphTransformer architecture to predict the semantic label of each joint node.
The model is optimized using a cross-entropy loss function (see more details in the Supp. Mat.).

\noindent\textbf{Semantic tokenization:} For a humanoid, we utilize the available semantic labels (\emph{e.g.}, main body, hair, cloth, accessories) to group bones by their semantics.
At the beginning of each group, we insert a special \btoken token to indicate the group’s start.
Then, within each group, we perform the DFS algorithm to preserve local topological consistency.
When selecting group roots, the main group takes the provided root node, while other groups choose the node closest or directly adjacent to the main group to maintain structural coherence. 
For intra-group traversal, child nodes are sorted by their spatial coordinates in $(z, y, x)$ order, ensuring alignment between topological and spatial hierarchies.
Finally, tokens are generated following a fixed group order (in practice, we arrange groups as main → hair → cloth → other).
Each group begins with its \btoken token, followed by the DFS-ordered bone tokens.
Each joint is represented as $6$ tokens, consisting of the discretized representation of its 3D coordinates and its corresponding parent 3D coordinates.
This design explicitly segments the skeleton into semantic groups while maintaining a stable spatial ordering and consistent parent–child encoding, making it suitable for auto-regressive sequence modeling.

Similarly, for tetrapods, we divide the skeleton into two categories: main bones and auxiliary bones.
Finally, we incorporate the corresponding positional embeddings to the obtained tokens, forming the skeleton token sequence \tboken.

\begin{figure}[h!]
  \centering
  \includegraphics[width=\linewidth]{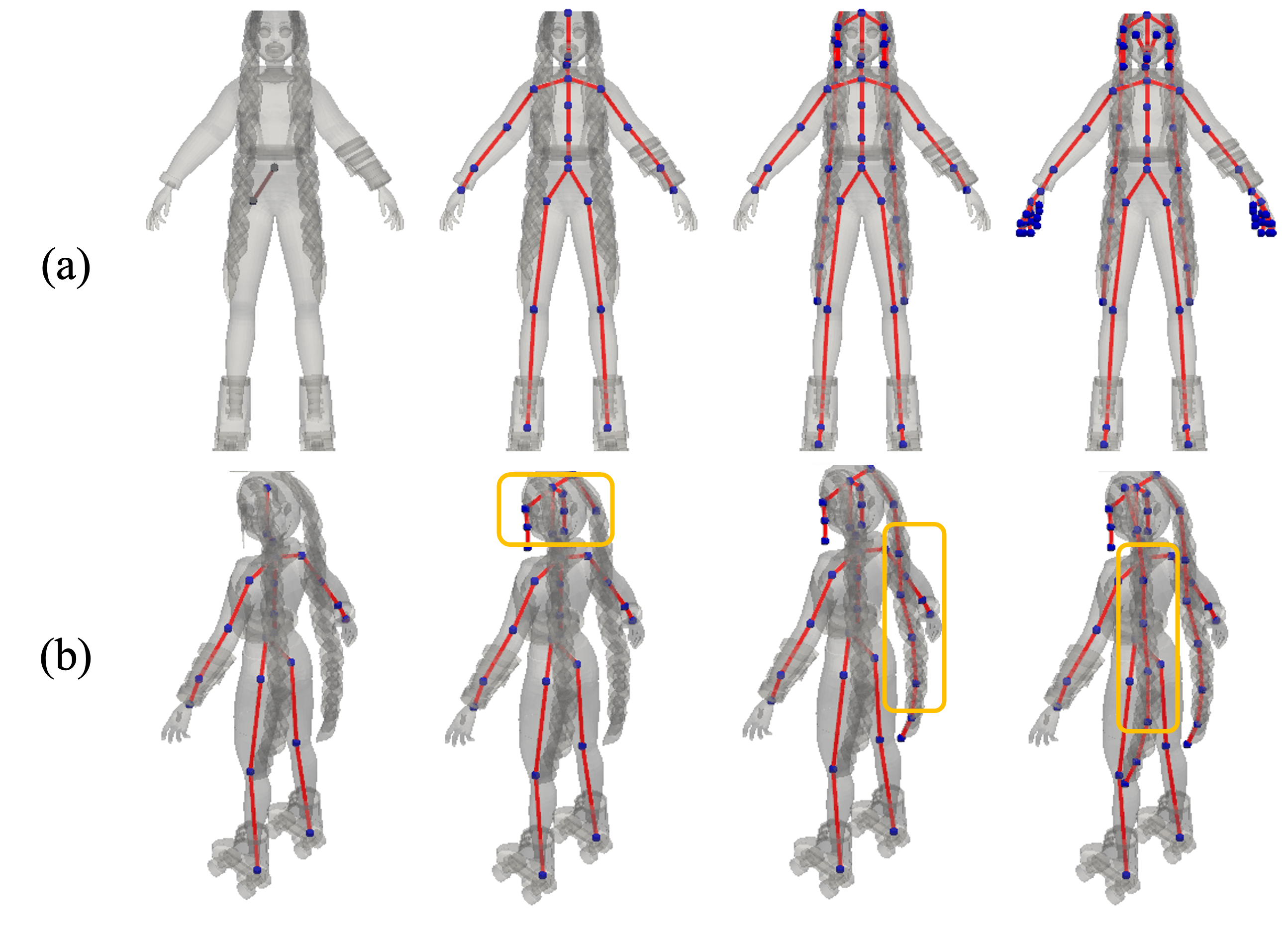}
    \caption{
    Our semantic-based skeleton tokenization. (a) illustrates our global skeleton grouping based on semantic categories, while (b) shows the within-group ordering following a DFS traversal.}
  \label{fig:tokenization}
\end{figure}

\noindent\textbf{Main Bones Control: }In practical industrial applications, the main bones are usually predefined and remain fixed to facilitate animation production and rigging pipelines. 
Animators typically build auxiliary bones (\emph{e.g., }clothing, hair) on top of the main bones (\emph{e.g., }main body) to achieve more detailed and flexible motion control. 
As mentioned in Sec.~\ref{sec:intro}, it is desirable that SG model can respect the coarsely crafted skeleton from animators and generate auxiliary bones on top of it. 

Surprisingly, our semantic-aware tokenization scheme can be used to realize such goal. 
More concretely, we consider this as a conditional generation task~\footnote{Certainly, our model can support unconditional generation.}. 
We tokenize the given main bones based on the semantic-based tokenization and compute their embeddings, which are concatenated with other conditioning vectors before decoding.
Thanks to this semantic-based tokenization design, the model learns to first generate the main-bone tokens followed by the auxiliary ones during training.
Consequently, by providing main-bone embeddings as conditional input, our auto-regressive decoder can seamlessly generate the corresponding auxiliary bones in a controllable and coherent manner.



\subsection{Learnable Density Control}\label{sec:LDT}

For different animation requirements, animators often need to design skeletons with varying numbers of bones for the same mesh, particularly by adjusting the number of auxiliary bones to achieve motions of different complexity. 
This motivates the need for controllable skeleton generation, where our model can regulate the number of generated bones to satisfy diverse animation design requirements.

Inspired by~\cite{tang2024edgerunner}, to enable flexible control over the number of generated bones, we introduce a learnable bone \ptoken token in the conditional input. 
Initially, we divide the bone count into discrete intervals, each represented by a control token. 
However, fixed interval thresholds fail to capture the continuous transition from simple skeletons (with only main bones) to complex ones (rich in auxiliary bones). 
To address this, we propose a \textbf{Learnable Density Interval} module that dynamically learns the interval cutpoints during training.

Specifically, we control the bone count via learnable binning with $K$ intervals. 
The global left/right edges are denoted as $e_0$ and $e_K$ (constants, non-trainable).
We define the learnable cutpoints as $\{c_i\}_{i=1}^{K-1}$ with monotonicity $c_1 < \cdots < c_{K-1}$.
We enforce monotonicity via cumulative softplus:
\begin{equation}
c_i = c_{i-1} + \text{softplus}(\Delta_i), \quad i = 2, \ldots, K-1.
\end{equation}
So the $K$ bins can be defined as:
\[
[e_0, c_1], (c_1, c_2], \ldots, (c_{K-2}, c_{K-1}], (c_{K-1}, e_K].
\]
For uniform notation, the left and right boundaries of bin $k$ are
\[
e_k^{\text{left}} =
\begin{cases}
e_0, & k = 1, \\
c_{k-1}, & k \ge 2,
\end{cases}
\qquad
e_k^{\text{right}} =
\begin{cases}
c_1, & k = 1, \\
c_k, & 1 \le k \le K-1, \\
e_K, & k = K.
\end{cases}
\]
We compute the soft bin probabilities as follows.
Given a bone count $n$ and a temperature parameter $\tau > 0$,
the soft probability of assigning $n$ to the $k$-th bin is defined as
\begin{equation} p_k(n) = \sigma\!\left(\frac{n - e_k^{\text{left}}}{\tau}\right) - \sigma\!\left(\frac{n - e_k^{\text{right}}}{\tau}\right), \quad k = 1, \ldots, K, 
\end{equation}
where $\sigma(\cdot)$ denotes the sigmoid function.
The probabilities are then normalized to ensure
$\sum_{k=1}^{K} p_k(n) = 1$.

Each bin is associated with a learnable embedding vector
$\mathbf{e}_k \in \mathbb{R}^C$ (distinct from the numeric boundary $e_k$).
The final bone-density conditioning vector is obtained as a probability-weighted combination:
\begin{equation} \mathbf{F}_{\text{density}}(n) = \sum_{k=1}^{K} p_k(n)\, \mathbf{e}_k, 
\end{equation}
with an optional hard inference mode using a one-hot selection at
$\arg\max_k p_k(n)$.

This learnable interval formulation allows the model to capture the underlying distribution of bone counts during training adaptively.
During inference, the learned cutpoints are fixed, providing stable, interpretable control over the generated skeleton's complexity.

\subsection{Full Model}\label{sec:gpt}


We now describe the auto-regressive generation process. 
We first uniformly sample $8,192$ points on the input mesh and compute their corresponding normal vectors. 
A pretrained point cloud encoder~\cite{zhao2023michelangelo} is then employed to extract the shape feature $\mathbf{F}_\text{shape}$, which serves as the conditional input to the transformer network.
To help the model distinguish whether auxiliary bones should be generated,
We divide the data into three types: humanoids with only main bones, humanoids with auxiliary bones, and non-humanoid shapes. 
A learnable classification token \clstoken is introduced and added as part of the conditional input.
We employ a decoder-only autoregressive model based on OPT-350M~\cite{zhang2022opt} to predict the sequence of discretized skeleton tokens \ttoken. 
To better inject the conditioning information, we not only prepend the conditional features before the \bostoken token as decoder inputs, but also insert a cross-attention layer after each self-attention layer,
where the hidden embeddings serve as the query and the conditional features as the key and value, allowing deeper fusion of conditional representations.
The network is trained using standard cross-entropy loss ($\mathcal{L}_{ce}$) to optimize token-level autoregressive prediction.

\begin{equation}\label{eqn:5}
\begin{aligned}
& \mathcal{L}_{ce} = CE(\hat{\mathbf{T}},  \mathbf{T}).
\end{aligned}
\end{equation}

\section{Experiments}\label{sec:exp}

\begin{table*}[!t]
\begin{tabular}{lclllllllll}
\hline
\rowcolor[HTML]{FFFFFF} 
                & Mode                                               & \multicolumn{1}{c}{\cellcolor[HTML]{FFFFFF}Precision $\uparrow$} & \multicolumn{1}{c}{\cellcolor[HTML]{FFFFFF}Recall $\uparrow$} & \multicolumn{1}{c}{\cellcolor[HTML]{FFFFFF}Accuracy $\uparrow$} & \multicolumn{1}{c}{\cellcolor[HTML]{FFFFFF}F1\_Score $\uparrow$} & \multicolumn{1}{c}{\cellcolor[HTML]{FFFFFF}J2J $\downarrow$}   & \multicolumn{1}{c}{\cellcolor[HTML]{FFFFFF}J2B $\downarrow$}   & \multicolumn{1}{c}{\cellcolor[HTML]{FFFFFF}B2B $\downarrow$} \\ \hline
\rowcolor[HTML]{FFFFFF} 
UniRig~\cite{zhang2025one}          & \cellcolor[HTML]{FFFFFF}                                                     & 0.105                               & 0.066                                & 0.078                               &0.077                                                     & \multicolumn{1}{c}{\cellcolor[HTML]{FFFFFF}0.038} & \multicolumn{1}{c}{\cellcolor[HTML]{FFFFFF}0.031} & \multicolumn{1}{c}{\cellcolor[HTML]{FFFFFF}0.026}                                                    \\
\rowcolor[HTML]{FFFFFF} 
Puppeteer~\cite{song2025puppeteer}       & \multirow{-2}{*}{\cellcolor[HTML]{FFFFFF}wo.train}                                   &  0.168                     &    0.086                  &   0.106                               & 0.105                                                & \multicolumn{1}{c}{\cellcolor[HTML]{FFFFFF}0.046} & \multicolumn{1}{c}{\cellcolor[HTML]{FFFFFF}0.038} & \multicolumn{1}{c}{\cellcolor[HTML]{FFFFFF}0.033}                                                      \\ \hline
\rowcolor[HTML]{FFFFFF} 
MagicArticulate~\cite{song2025magicarticulate} & \multirow{-1}{*}{\cellcolor[HTML]{FFFFFF}retrain}                                        &              0.712                                         &     0.701                                               &  0.697                                               &   0.707                                                  &                                        0.044           &                                            0.034       &                                                0.032                                                   \\
\rowcolor[HTML]{E7E6E6} 
Ours            & \multicolumn{1}{l}{\cellcolor[HTML]{E7E6E6}}                                                     &     \textbf{0.745}                                                  &                                                 \textbf{0.731}   &                                               \textbf{0.729}  &                                                  \textbf{0.730}   &                                                \textbf{0.036}   &                                                \textbf{0.027}   &                                                  \textbf{0.025}                                                      \\ \hline
\end{tabular}
\caption{
Joint prediction results on the test set. MagicArticulate is retrained on our dataset.
}\label{table:joints}
\end{table*}

\subsection{Implementation Details}
To enhance the model’s robustness and generalization ability, we apply geometric data augmentations, including scaling, translation, and rotation transformations.
All experiments are conducted with a batch size of $12$.

\subsection{Metircs and Baselines}
\textbf{Baselines: }We compare our method against three representative baselines for automatic skeleton generation: UnRig~\cite{zhang2025one}, which incorporates template prompting into auto-regressive generation. 
Puppeteer~\cite{song2025puppeteer}, an auto-regressive framework that improves skeleton connectivity. 
MagicArticulate~\cite{song2025magicarticulate}, which uses an auto-regressive transformer approach.  
All methods are evaluated on our test dataset. 

\noindent\textbf{Metrics: }We evaluate skeleton generation quality using eight metrics. 
To evaluate whether different methods can generate sufficiently detailed skeletal structures of objects, we calculate Precision, Recall, Accuracy, and F1-Score by comparing predicted and ground-truth joint positions within a spatial distance threshold $\tau$. A prediction is considered correct if its Euclidean distance to any ground-truth joint is below $\tau$. We further employ three Chamfer–Distance–based metrics —CD-J2J (joint-to-joint), CD-J2B (joint-to-bone), and CD-B2B (bone-to-bone) —that measure spatial alignment between predicted and ground-truth skeletons.
$\tau$ is set to 0.01 in our experiments.

\begin{figure*}[!t]
  \centering
  \includegraphics[width=\linewidth]{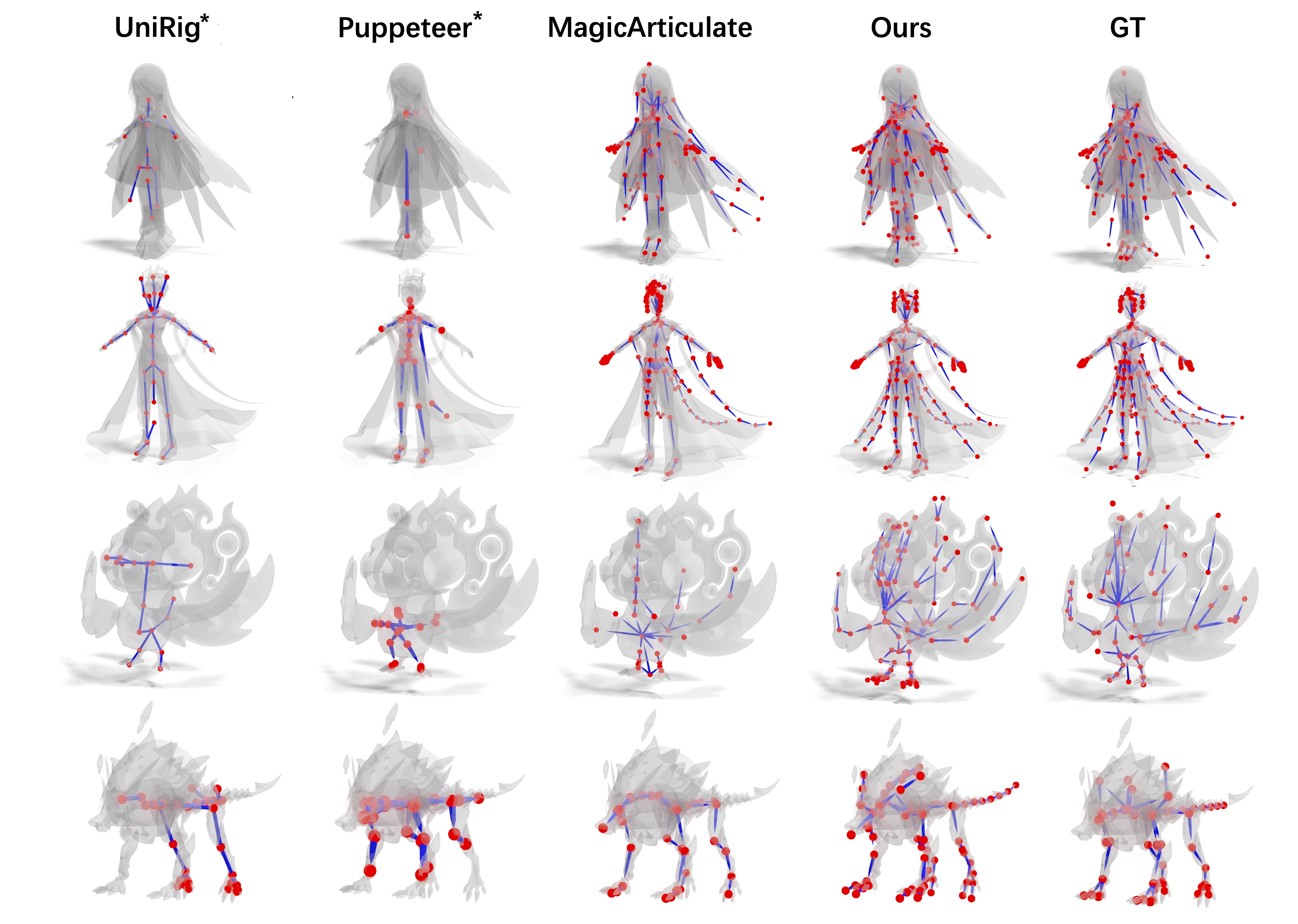}
    \caption{
   Comparison of skeleton generation results on our test set, and \(*\) indicates the method is directly inferred with a publicly available checkpoint. 
   Our method produces skeletons that are more detailed and structurally complete.}
  \label{fig:baseline}
\end{figure*}

\subsection{Evaluation}
\noindent\textbf{Quantitative comparison:} The quantitative results are shown in Table ~\ref{table:joints}. We retrain the MagicArticulate ~\cite{song2025magicarticulate} model on our dataset. 
Our method consistently outperforms baselines across nearly all metrics. 
In particular, compared with UnRig and Puppeteer, our model shows a five- to nine-fold increase in Precision and F1-Score, demonstrating that our model can generate more complete and fine-grained skeletal structures. Even when compared with the retrained MagicArticulate model, our model still maintains strong competitiveness.
Since the first two baselines were never trained on our dataset, they can only predict overly simplified skeletal structures. Although MagicArticulate is trained on our data, it still lacks an understanding of complex skeletal topology and therefore fails to generate high-quality auxiliary bones.

\noindent\textbf{Qualitative comparison:} Qualitative comparisons are presented in Figure ~\ref{fig:baseline}. 
UniRig and Puppeteer tend to produce sparse or incomplete skeletons, often missing fine-grained joints such as hands, tails, or hair accessories.  As shown in Figure~\ref {fig:baseline} (second row, first and second column), the clothing-related skeletal structures are completely missing in the predictions.
Retrained MagicArticulate improves the overall joint coverage but still suffers from inaccurate topology and misplaced bones in complex structures.  As shown in Figure~\ref {fig:baseline} (second row, third column), the head-region skeletons are severely disorganized, with joints misplaced or entangled in the surrounding mesh.
In contrast, our method generates structurally complete skeletons, closely aligning with the ground truth across diverse categories.

\subsection{Ablation study}
We design two types of control tokens in the model’s conditional input: the Density token to control the density of the generated skeleton, and the CLS token to indicate the input category.
Experimental results show that introducing these tokens not only enables controllable generation but also improves prediction performance, see the upper part of Tab.~\ref{table:joints_abl}.
Specifically, adding the density token reduces the joint-to-joint (J2J) distance by 12.2\%.
We also conduct ablation studies on skeleton tokenization strategies, comparing (1) a naive global DFS-based tokenization and (2) a semantic grouping method without local DFS ordering.
The results are shown in the lower part of Tab.~\ref{table:joints_abl}, compared with these two baselines, our proposed semantic-based tokenization achieves 16.3\% and 10\% lower J2J distances, respectively, demonstrating its effectiveness.

\begin{table}[h!]
\begin{tabular}{lcccc}
\hline
\rowcolor[HTML]{FFFFFF} 
                    & Accuracy $\uparrow$ & J2J $\downarrow$   & J2B $\downarrow$   & B2B $\downarrow$   \\ \hline
\rowcolor[HTML]{FFFFFF} 
wo. Density   token & 0.699    & 0.041 & 0.033 & 0.032 \\
\rowcolor[HTML]{FFFFFF} 
wo. CLS token       & 0.714    & 0.037 & 0.028 & 0.027 \\ \hline
\rowcolor[HTML]{FFFFFF} 
naive tokenization  & 0.701    & 0.043 & 0.032 & 0.031 \\
\rowcolor[HTML]{FFFFFF} 
wo. Part DFS        & 0.712    & 0.040 & 0.029 & 0.028 \\
\rowcolor[HTML]{FFFFFF} 
w. Part BFS        & 0.723    & 0.037 & 0.028 & 0.026 \\ \hline
\rowcolor[HTML]{E7E6E6} 
Full Model                & \textbf{0.729}    & \textbf{0.036} & \textbf{0.027} & \textbf{0.025} \\ \hline
\end{tabular}
\caption{
Ablation study on the test set.
}\label{table:joints_abl}
\end{table}

\subsection{Applications}
In this section, we demonstrate two practical applications of our method: controlling the density of generated skeletons and generating auxiliary bones conditioned on given main bones.
Both applications address key practical demands in real-world rigging scenarios, and to the best of our knowledge, we are the first to explore these directions.
\begin{figure}[b!]
  \centering
  \includegraphics[width=\linewidth]{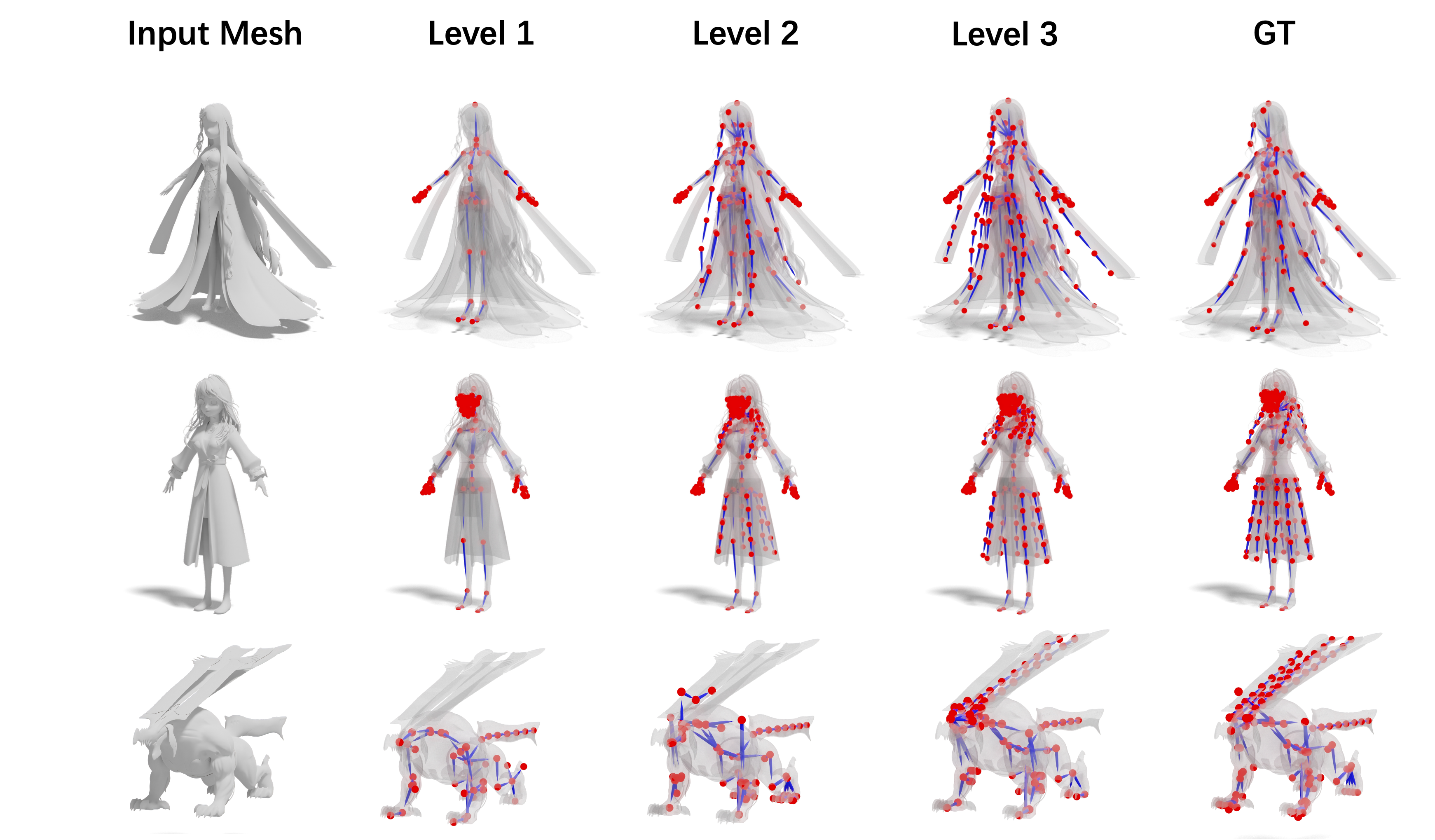}
    \caption{
   Density control results. Our model enables the generation of skeletons with controllable density.}
  \label{fig:density_cond}
\end{figure}

\noindent\textbf{Density control: }During training, we introduce the learnable density token that enables the model to capture the distribution of skeletons with varying densities from large-scale data.
We initialize three density levels according to the empirical distribution of main and auxiliary bones — [0–50], [50–150], and $>$150 — corresponding to low (Level 1), medium (Level 2), and high (Level 3) density, respectively.
At inference time, we can control the sparsity of the generated skeletons by adjusting the density token. As shown in Fig.~\ref{fig:density_cond}, increasing the density token generally leads to more complex and plausible skeleton structures.
Remarkably, our model successfully learns the structural priors of skeleton distribution from data: when increasing the density token, the main bones remain stable, while more reasonable auxiliary bones are generated.
For humanoid characters, the model tends to add bones around skirts, ribbons, and accessories, extending bone chains naturally.
For non-humanoid objects, it adds bones around non-torso regions and attached parts, aligning well with real-world rigging requirements.

\begin{figure}[t!]
  \centering
  \includegraphics[width=\linewidth]{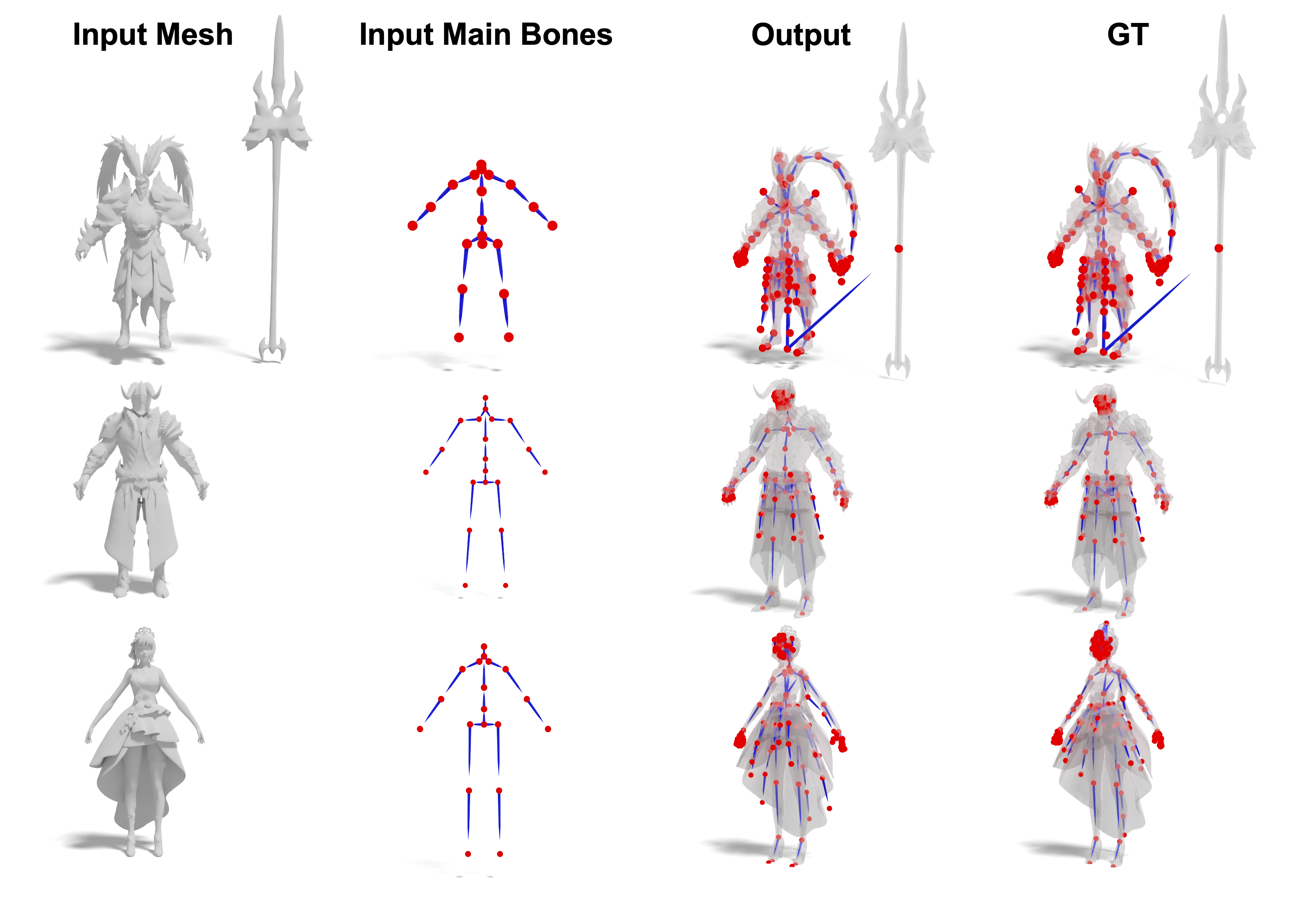}
    \caption{
   Main bones control results. Our model can generate the fine-grained auxiliary bones conditioned on the main bones.}
  \label{fig:main_cond}
\end{figure}

\noindent\textbf{Main bones control: }In practical scenarios, the main bones of characters are often generated from predefined templates and thus cannot be modified, while the auxiliary bones need to be created on top of them — a process that is typically time-consuming and labor-intensive.
Thanks to our proposed semantic-based tokenization scheme, where main and auxiliary bones are explicitly grouped and ordered, our model enables flexible conditional generation during inference: given the template main bones, it can generate the corresponding auxiliary bones automatically, which is difficult to achieve with naive tokenization methods.
As shown in Fig.~\ref{fig:main_cond}, our approach can produce complex and accurate auxiliary structures conditioned on the given meshes and main bones, including skirts, hair strands, and accessories, demonstrating strong generalization and controllability.

\section{Conclusion, Limitation, and Future Work}\label{sec:conclusion}
In this work, we present the first controllable skeleton generation framework capable of producing high-complexity, fine-grained, and semantically consistent skeletons. 
By constructing a large-scale rigging dataset of $82,633$ rigged meshes and introducing a semantic-aware tokenization scheme, our model learns both structural topology and part-level semantics in a unified manner. 
The proposed density token and main-bone–conditioned generation enable explicit structural controllability. 
We believe this work takes an important step toward practical and intelligent rigging automation.

We also identify the following limitations. 1) Although our dataset covers a wide range of articulated forms, certain categories—such as vehicles and accessories are underrepresented, limiting the model’s generalization in these domains; 2) while our density token enables global bone-density control, it does not yet support precise local control over bone counts within specific regions. 

Future research will focus on developing finer-grained mechanisms for region-level skeletal density control. Another promising direction is the integration of fully automated animation generation on top of the produced skeletons, which would further advance the rigging and animation pipeline.

\section*{Acknowledgments}
This work is supported in part by the National Natural Science Foundation of China under contract No. 62171256 and Meituan, in part by the Tencent VISVISE team.

{
    \small
    \bibliographystyle{ieeenat_fullname}
    \bibliography{main}
}


\end{document}